\documentstyle[epsfig,float,twocolumn,prb,aps]{revtex}
\newcommand{\figwidth}{6.0cm} 
\newcommand{\figheight}{8.2cm} 

\begin{document}

\draft 
\twocolumn[\hsize\textwidth\columnwidth\hsize\csname @twocolumnfalse\endcsname

\title{
Dynamical Correlation Functions using the Density Matrix Renormalization Group
} 

\author{Till~D.~K\"uhner$^{1,2}$ and Steven~R.~White$^1$} 

\address{
$^1$Department of Physics and Astronomy,University of California,Irvine, CA 92697\\
$^2$Physikalisches Institut der Universit\"at Bonn,
  D-53115 Bonn,   Germany\\
}

\date{\today}  \maketitle 

\widetext

\begin{abstract}
{
  \noindent
{
The density matrix renormalization group (DMRG) method allows for very
precise calculations of ground state properties in low-dimensional 
strongly correlated systems.
We investigate two methods 
to expand the DMRG to calculations of dynamical properties.
In the Lanczos vector method the DMRG basis is optimized to represent
Lanczos vectors, which are then used to calculate the spectra.
This method is fast and relatively easy to implement, but the accuracy 
at higher
frequencies is limited.
Alternatively, one can optimize the basis to represent a correction vector
for a particular frequency. The correction vectors can be used to
calculate the dynamical correlation functions at these frequencies with high 
accuracy. By separately calculating correction vectors at different
frequencies, the dynamical correlation functions can be interpolated and
pieced together from these results.
For systems with open boundaries we discuss how to construct operators
for specific wavevectors using filter functions.
}
}
\vspace{0.3in}
\end{abstract}


]

\narrowtext


\section{Introduction}

Since its development, the density matrix 
renormalization group\cite{White:1992,White:1993} (DMRG) has
been successfully used to calculate static properties of ground states and
low-lying excited states in various low dimensional strongly interacting 
systems. Energies can be determined with highest precision, and the 
calculation of time-independent correlation functions is easy and 
high accuracy can be achieved. The calculation of dynamical properties
is more difficult.

The Lanczos vector method, also known as the continued fractions method,
can be used to determine
the dynamical correlation functions in an exact diagonalization 
calculation. However, in a DMRG calculation,
if the basis is optimized only to represent the ground state, this will
lead to poor results, since 
the Lanczos vectors are not represented correctly in the truncated basis.
Hallberg suggested using several of the first Lanczos vectors as target 
states in addition to the ground state\cite{Hallberg:1995}. 
Reasonable results were obtained for an S=1/2 chain, but the true accuracy
of the method was not determined, since the infinite system DMRG method,
rather than the finite system method, was used.
In this paper we determine the accuracy of this method for the more
appropriate finite system method.

An alternative approach for generating dynamical spectra is the correction
vector method, which yields exact results 
within the given basis\cite{SoosRamasesha:1989}. Although it has been used 
in the DMRG context\cite{Pati:1997}, in those
calculations the basis was not expanded to represent these correction
vectors. Here we apply the correction vector method, targeting a 
correction vector at a particular frequency.
We find that the dynamical correlation function at this frequency can
be calculated directly and very accurately. We show how to piece
together results from correction vectors at different frequencies to obtain
the full spectrum.

Taking antiferromagnetic Heisenberg chains with spin 1 
and spin $1/2$ as examples, we discuss the advantages and 
limitations of the two methods. We show how to obtain the spectral 
weight functions, and how to judge the quality of the numerical results.

DMRG calculations are most accurate with open boundary conditions, in
which case momentum is not precisely defined. In this work we 
show how to construct operators corresponding to wavevectors in systems
with open boundaries  using filter functions.

In Section \ref{Filtersection} we discuss the construction of the
operators for systems with open boundaries. In Section \ref{LanczosVecSect}
we present the Lanczos method, and in Section \ref{Spin1Sect} we 
apply it to the antiferromagnetic spin-1 chain. As an example where
the Lanczos method does not work so well we discuss the 
antiferromagnetic spin-1/2 chain in Section \ref{Spin1d2Sect}. In Section
\ref{CorrVecSect} we present the correction vector method, and we give
conclusions in Section \ref{ConclusionSect}.

\section{Construction of operators for open systems}
\label{Filtersection}

To calculate a Green's function
\begin{equation}
	G(q,z) =\langle 0 | A_q^{\dagger}\frac{1}{z-H} A_q | 0 \rangle
	\label{GreensFunction}
\end{equation}
with $z=\omega+ i \eta$, we have to be able to apply an operator $A_q$ in our 
system. 
The DMRG works in real-space, and operators with wavevectors $q$ can be
obtained as Fourier transforms of on-site operators $A_n$. 
In infinite systems they are given as:
\begin{equation}
	A_q = \sum_{n=-\infty}^{\infty} A_n e^{i x_n q} \;,
\label{FourierTransformEq}
\end{equation}
\begin{equation}
	A_n  = \frac{1}{2\pi} \int_{-\pi}^{\pi} dq A_q  e^{-i x_n q} \;.
\end{equation}
where $x_n$ is the position of site $n$, the lattice spacing is $a=1$.

For finite systems with open boundaries, we construct operators defined 
as wavepackets,
with finite spatial extent and finite uncertainty in the momentum.
We construct the wavepacket by inserting a windowing or filter
function in Eq.(\ref{FourierTransformEq}).
If only real operators are used, it is numerically convenient to
construct
\begin{eqnarray}
	\nonumber
	A(q) &=& \sum_{n=-\infty}^{\infty} \sin(q x_n) f(x_n) A_n\\
	&=& \frac{1}{4 \pi i} \int dq' (A_{q'} - A_{-q'}) f(q-q') \;,
	\label{sinappl}
\end{eqnarray}
and
\begin{eqnarray}
	\nonumber
	A(q)& =& \sum_{n=-\infty}^\infty \cos(q x_n) f(x_n) A_n\\
	&=&  \frac{1}{4 \pi}  \int dq' (A_{q'} + A_{-q'}) f(q-q')
	\label{cosappl}
\end{eqnarray}
where $f(x_n)=F(\frac{x_n}{M})$ is the filter function, $2 M$ is the width 
of the window. We use system with even numbers of sites, 
and $x_n$ is offset so that \mbox{$x=0$} is in the center 
of the system. The sites closest to the middle of the system
are at $x=-1/2$ and $x=1/2$.
The operators $A(q)$ are reflection symmetric, which allows, 
if the Hamiltonian is also reflection symmetric,  
using reflected system blocks as environment blocks in the DMRG.

Applying the operator as if the system were periodic, is equivalent 
to using a rectangular window $F^r(x)$ as the filter function with $2 M = L$:
\begin{equation}
	F^r(x) = \left\{
	\begin{array}{lll}	
		1 & \text{if} & -1 \leq x \leq 1 \\
		0 &  & \text{otherwise}
	\end{array}
	\right. 
	\, 
\end{equation}

This operator is seriously flawed. First, it has substantial
weight at the edges of the system, where open boundary effects are significant.
Second, even if we ignore the edge effects, this window is very broad in
the wavevector space. The Fourier transform of $F^r(x)$ is
\begin{eqnarray}
	F^r(q) 	=& \sin(\frac{L q}{2}) \left( \sin(\frac{q}{2}) 
			- \frac{\sin(q) \cos(\frac{q}{2})}{\cos(q)-1}
			\right) \;,
\end{eqnarray}
which for small $q$ is
\begin{eqnarray}
	F^r(q) 	\approx& \frac{2 \sin(L/2 q)}{q} \,.
\end{eqnarray}

The wavevector uncertainty $\Delta q$ of this operator is of order 1, even
when $L \rightarrow \infty$, whereas $q$ itself ranges from 0 to $2\pi$.
Therefore this operator is not useful.

\begin{figure}[t]
  \begin{center}
    \epsfig{file={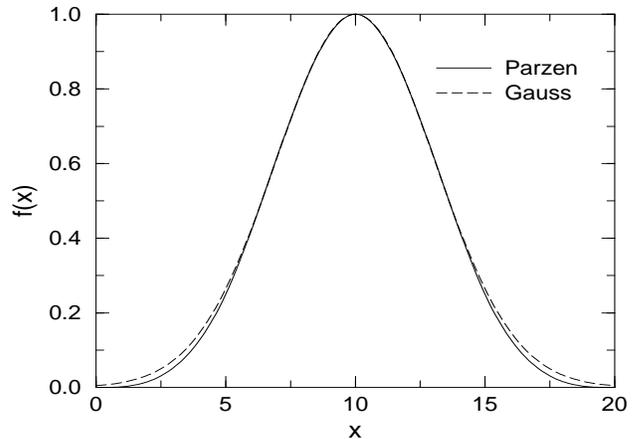}, height=\figheight,width=\figwidth, angle=-90}
    \caption
    {
	Parzen and Gaussian functions in a $L=20$ site system. The 
	halfwidth of the Parzen function is approximately $0.18 L$. 
	The Gaussian has a standard deviation of 
	$\sigma \approx 0.153 L$, while the standard deviation of the 
	Parzen window, because of the faster decay of its tails,
	is only $\sigma \approx 0.112 L$.
   }
	\label{Parzen_Filter.plot}
  \end{center}
\end{figure}

As is well known from elementary quantum mechanics, the wavepacket with
the minimum product of uncertainties ${\Delta q} {\Delta x}$ is a Gaussian
function. 
However, it is more desirable to use an operator which is equal to 0 at the 
edges of the system. A widely used filter that looks similar to a Gaussian
in the center, but with compact support, is the Parzen 
filter (Fig. \ref{Parzen_Filter.plot}):
\begin{equation}
\label{Parzen_filter}
	F_p(x) = \left\{
	\begin{array}{lll}	
		1- 6 |x|^2 + 6 |x|^3 
			& \text{if} & 0\leq |x| \leq 1/2 \\
		2 (1-|x|)^3 & \text{if} & 1/2 \leq |x| \leq 1 \; .
	\end{array}
	\right. 
\end{equation}

\begin{figure}[t]
  \begin{center}
    \epsfig{file={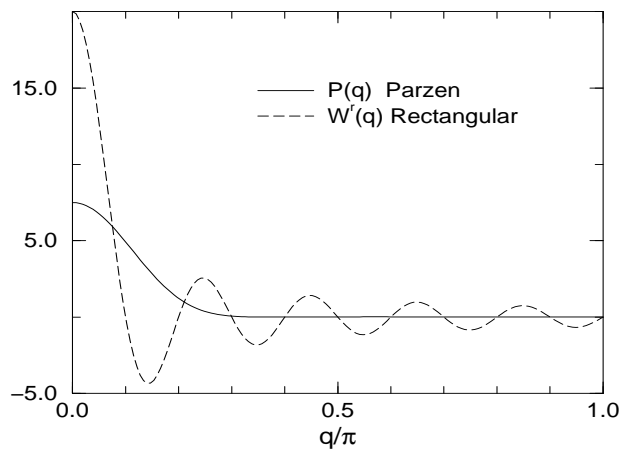}, height=\figheight,width=\figwidth, angle=-90}
    \caption
    {
	The Fourier transform of the Parzen filter and the rectangular window.
	In real space both filters are 20 sites wide. It is obvious that the
	Parzen filter has a smaller width in wavector space.
    }
	\label{Parzen_four.plot}
  \end{center}
\end{figure}

With $2  M = L$ this filter smoothly goes to zero at the boundaries. 
The Fourier transform is
\begin{eqnarray}
	F_p(q) 	\approx& 24 \frac{3 + \cos(q M) - 4 \cos(q M/2)}{q^4 M^3}
\end{eqnarray}
and the wavevector uncertainty is $\Delta q = 2 \sqrt{3}/M$. 
We see that $\Delta q$ varies inversely with the system size
if $2 M =L$.
For example, in a system with 100 sites 
$\sigma \approx 0.07$, 
and with wavevectors $0 \leq q \leq 2 \pi$ this is a relative error of only 
$\Delta q = 1.1\%$. Figure \ref{Parzen_four.plot} shows how the Fourier 
transform of the Parzen filter smoothly goes to zero, while the Fourier 
transform of a rectangular window has oscillations at higher frequencies.

The use of  sine (Eq. \ref{sinappl}) and cosine (Eq. \ref{cosappl}) 
functions for the operators instead of the 
complex form (Eq. \ref{FourierTransformEq}) requires some special
attention. The Green's function is 
\begin{eqnarray}
	\nonumber
	G(q,z)=&&\langle A^{\dagger}(q) \frac{1}{z-H} A(q) \rangle \\
	=&&\frac{1}{4} \langle (A^{\dagger}_{q}\pm A^{\dagger}_{-q} )
			\frac{1}{z-H} (A_{q} \pm A_{-q}) \rangle \;.
\end{eqnarray}
with ``$-$'' for sine and ``$+$'' for cosine. 
Noting that $\langle A^{\dagger}_{-q} \frac{1}{z-H} A_{-q} \rangle
= \langle A^{\dagger}_{q} \frac{1}{z-H} A_{q} \rangle$
and $A_{-\pi} = - A_{\pi}$, there are three different cases  
for the Green's function. For \mbox{$A(q)=\sum_n \cos(q n) A_n $}
it is:
\begin{eqnarray}
	G(q,z)=&&	
	 \left\{ 
	\begin{array}{cl}
		\frac{1}{2} \langle A^{\dagger}_{q} 
			\frac{1}{z-H} A_{q} \rangle 
			&  \text{for $0<q<\pi$} \\
		 \langle A^{\dagger}_{q} 
			\frac{1}{z-H} A_{q} \rangle 
			&  \text{for $q=0$} \\
		0	&  \text{for $q=\pi$} \\
	\end{array} \right.
\end{eqnarray}

For \mbox{$A(q)=\sum_n \sin(q n) A_n $}:
\begin{eqnarray}
	G(q,z)=&&	
	 \left\{ 
	\begin{array}{cl}
		\frac{1}{2} \langle A^{\dagger}_{q} 
			\frac{1}{z-H} A_{q} \rangle 
			&  \text{for $0<q<\pi$} \\
		0	&  \text{for $q=0$} \\
		 \langle A^{\dagger}_{q} 
			\frac{1}{z-H} A_{q} \rangle 
			&  \text{for $q=\pi$} \\
	\end{array} \right.
\end{eqnarray}

If only sine or cosine functions are used, the values found at $q=0$
and $q=\pi$ are either zero or twice the expected values. By doing 
separate calculations with sine and cosine functions and adding up 
the results the correct values are always obtained.

In finite systems this effect is broadened due to the  
finite width $\Delta q$ of the operators in the wavevector space. 
To get system size independent values for the Green's functions
we require $\pi^2 = \int dq  f(q)^2$ for the filter functions.  
For the Parzen filter this means that an additional prefactor
of $\sqrt{\frac{140 \pi}{151 M}}$ must be included in the filter.

\section{Lanczos vector method}
\label{LanczosVecSect}

To calculate spectral functions
a Lanczos vector procedure can be used\cite{Hallberg:1995}. To do this,
the Hamiltonian $H$ 
is projected onto a Krylov subspace spanned by Lanczos vectors $|f_n\rangle$:
\begin{eqnarray}
	\nonumber
	|f_0\rangle =&& A_q |0\rangle  / \langle0|  A_q^{\dagger} A_q |0\rangle\\
\nonumber
	a_n  	=&& \langle f_n|H|f_n\rangle \\
\nonumber
	b_n  	=&& \|{|r_n\rangle}\|_2 \\
\nonumber
	|r_n\rangle =&& (H-a_n) |f_n\rangle 
			- b_{n-1} |f_{n-1}\rangle \\
	|f_{n+1}\rangle =&& |r_n\rangle/b_n
\end{eqnarray}

In the Lanczos basis the Hamiltonian is tridiagonal:
\begin{eqnarray}
\nonumber
	H = \left(
	\begin{array}{cccccc}
		a_0    & b_0    &         &         & &0 \\
		b_0    & a_1    & b_1     &         &  \\
		       & b_1    & a_2     & b_2     &  \\
		       &        & \ddots & \ddots  & \ddots  &  \\
		 \phantom{b_{n-1}}&    &     & b_{n-2} & a_{n-1} & b_{n-1} \\
		0 & \phantom{b_{n-1}}       &  \phantom{b_{n-1}}      &  \phantom{b_{n-1}}       & b_{n-1} & a_n 
	\end{array}	
	\right)	
\end{eqnarray}

Now the eigenvectors $|\Phi_n\rangle$ of $H$ are used as an approximation
of the identity $1 \approx \sum_n |\Phi_n\rangle \langle \Phi_n |$. 
Inserting this into the Green's function we get:
\begin{eqnarray}
	\nonumber
	G(q,z) &\approx & \sum_{n,m}
	  \langle 0 | A_q^{\dagger} |\Phi_n\rangle \langle \Phi_n | 
\frac{1}{z-H} |\Phi_m\rangle \langle \Phi_m | A_q | 0 \rangle \\
	\nonumber &= &
	\sum_n  \langle \Phi_n |  \frac{1}{z-H} |\Phi_n \rangle
	 {\langle \Phi_n | A_q | 0 \rangle}^2 \\
	&= &
	\sum_n  \frac{(\Phi_n^0)^2  \langle 0 | A_q^\dagger A_q | 0 \rangle}{z-E_n}  
\end{eqnarray}
Here $E_n$ is the eigenvalue of $\Phi_n$, and $\Phi_n^m = \langle f_n |\Phi_n\rangle$ . 
The dynamical correlation function $I_A(q,\omega)$ is then given by
\begin{eqnarray}
	\nonumber 
	I_A(q,\omega) 
    &=& - \frac{1}{\pi} \text{Im} \lim_{\eta \rightarrow 0^+} 
	 G(q,\omega + i \eta + E_g)\\
	\nonumber
    &=& \frac{\langle 0 | A_q^\dagger A_q | 0 \rangle}{\pi} \lim_{\eta \rightarrow 0^+} 
	\sum_n  \frac{\eta \; (\Phi_n^0)^2}{(\omega+E_g-E_n)^2 + \eta^2} \\
    &=& \langle 0 | A_q^\dagger A_q | 0 \rangle \sum_n \delta(\omega+E_g-E_n) (\Phi_n^0)^2 \; ,
\end{eqnarray}
where $E_g$ is the ground state energy. The peaks in the correlation
function are at $\omega_n = E_n - E_g$.

The major source of errors in this is the approximation of the identity in
terms of Lanczos vectors, and the small number $N_L$ of these Lanczos vectors
that can be used as target states in practical calculations. Only those
states that are used as target states can be relied on as being represented
correctly. Similar to the Lanczos algorithm for the solution of eigensystems,
every additional Lanczos vector adds another peak, typically at 
high energy,
while peaks with the smallest frequencies are determined most precisely
and converge the fastest.

To obtain the dynamical correlation function at the end of the calculation, we
make full use of the available Hilbert space. We calculate $I(q,\omega)$
at the DMRG step where the system block is the same size as the environment 
block, since the truncation errors are smallest in that step.
To do this, we not
only use the Lanczos vectors that were target states, but
keep calculating new Lanczos vectors until orthogonality breaks down and
the overlap of the new Lanczos vector $| f_n \rangle$ with the 
first Lanczos vector $| f_0 \rangle$
is bigger than $1\%$.

If more than just a few Lanczos vectors are used, the question of the weight
that is assigned to these target states in the density matrix 
has to be addressed. 
The weight of a Lanczos vector in the spectrum is given by:
\begin{eqnarray}
	w_n = \sum_m (\Phi_m^0)^2 (\Phi_m^n)^2 \;.
\label{VectorWeight}
\end{eqnarray}
Taking this as a measure for the importance of a Lanczos vector,
we assign $50\%$ of the weight to the ground state, and distribute the 
remaining $50\%$ among the Lanczos target states according to their weight
$w_n$.

\section{The spin 1 Heisenberg model}
\label{Spin1Sect}
As an example we look at the antiferromagnetic spin 1 Heisenberg model:
\begin{equation}
	H = \sum_i \vec{S}_i \vec{S}_{i+1} \;.
\end{equation}
Each of the spins can be viewed as two spin-1/2 spins that pair with the 
spin-1/2 on the neighboring site in an antisymmetric singlet 
wavefunction\cite{AKLT}. In an open chain this effectively leaves 
unpaired spin-1/2's at the ends. To compensate for this, we add real 
spin-1/2's at the ends of the chain. We do not include
theses spins in the calculation of the spectra and set $2 M = L-2$ 
for the Parzen filter. 

This system has a finite correlation length \mbox{$\xi\approx 6.03(1)$} and 
the Haldane gap \mbox{$\Delta_H = 0.41050(2)$\cite{WhiteHuse},}
that separates the ground state from the first excitation, a single magnon
with wavevector $\pi$. DMRG is particularly accurate in this system
even if the number of states kept is small.

\begin{figure}[h]
  \begin{center}
    \epsfig{file={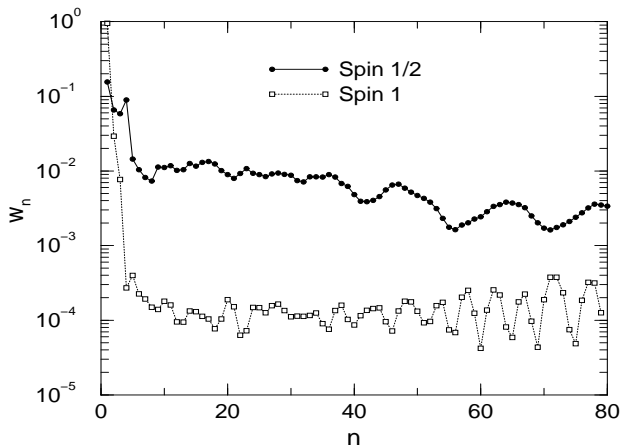}, height=\figheight,width=\figwidth, angle=-90}
    \caption
    {
	The weight $w_n$ of the Lanczos vectors in the spectrum. $L=320$,
	$N_L=3$ Lanczos target states, $m=128$ states for Spin 1 and 
	$m=256$ states for Spin 1/2, $q=\pi$. Weight for the first three
	vectors: $\sum_{n=0}^2 w_n = 0.2799$ 
	for Spin 1/2, and $\sum_{n=0}^2 w_n = 0.9887$ 
	for Spin 1.
    }
	\label{weight_vs_Lanczosvector.Spin1_and_1d2}
  \end{center}
\end{figure}

To obtain the dynamical correlation function we keep the ground state and the
three first Lanczos states as target states ($N_L=3$), 
and we keep $m=128$ states in the
DMRG basis. To verify that keeping three Lanczos vectors as target states
is enough, we calculate the weight of the Lanczos vectors in the final 
$S^+(q,\omega)$ for $q=\pi$. 
Figure \ref{weight_vs_Lanczosvector.Spin1_and_1d2} shows that the weight
of the first vectors is big and decays fast, and 
the first three Lanczos vectors contain $98.87\%$ of the total weight.

\newpage

\begin{figure}[h]
  \begin{center}
    \epsfig{file={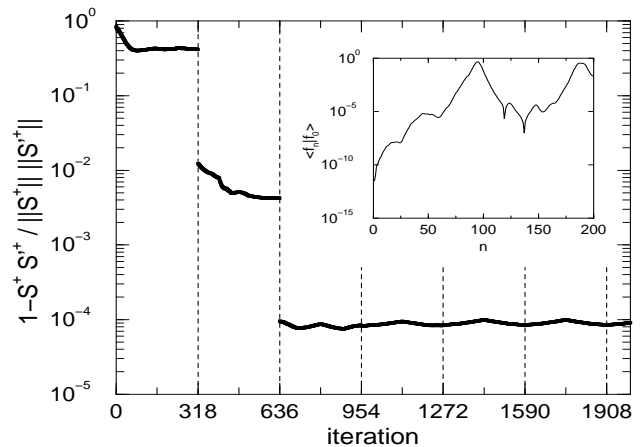}, height=\figheight,width=\figwidth, angle=-90}
    \caption
    {
	Deviation from unity of the overlap of the spectral weight calculated 
	in every step with the spectral weight calculated in the last
	step. Spectral weights at $q=\pi$ in a 320 site long
	spin-1 chain, with a broadening factor $\eta=0.01$ and $m=128$
	states kept.
	Iterations are counted from
 	the first step after the build-up phase. The dashed lines indicate
	the iterations at which the system and environment block have the same
	size. 
	The inset shows the overlap of the first Lanczos vector 
	$|f_0\rangle$ with $|f_n\rangle$ in the final calculation 
	of the correlation function. 
    }
	\label{SP1.int_final_spectrum_vs_iteration.L=320.eta=0.01.128states}
  \end{center}
\end{figure}

To further verify the convergence of the DMRG basis, we 
calculate $S'^+(q,\omega)$ in every step and compare it to the final 
result  $S^+(q,\omega)$ by taking the inner product of the two:
\begin{equation}
	\frac{S'^+(q) \cdot S^+(q)}{\left\| S'^+(q)\right\|  \left\| S^+(q)\right\|} \;,
\label{product.eq}
\end{equation}
with the inner product defined as 
$A \cdot B \equiv \int_{-\pi}^\pi d\omega A(\omega) B(\omega)$,
and the norm $\left\|A\right\| = \sqrt{A \cdot A}$. For this calculation we 
use a finite broadening factor $\eta=0.01$.
Figure \ref{SP1.int_final_spectrum_vs_iteration.L=320.eta=0.01.128states} shows
the product for every DMRG step. The discontinuities occur when the system 
and environment blocks have the same size, and utilizing reflection symmetry,
the system block is reflected onto the environment block.
After two sweeps the DMRG basis is converged, there are no further
discontinuities, only small oscillations which are due to the 
different truncation effect depending on the size of the system 
and the environment block. Obviously with three Lanczos target
states the spectrum is described well enough, and convergence is
very good.

In the calculation of the final spectrum, Lanczos vectors are calculated
until the overlap of the new vector with the first one exceeds $1\%$. 
The inset of 
Fig. \ref{SP1.int_final_spectrum_vs_iteration.L=320.eta=0.01.128states}
shows that the overlap $\langle f_n | f_0 \rangle$  increases exponentially. 
In the given example we use the first 83 Lanczos vectors. It can be seen from
the overlap that orthogonality breaks down after 95 Lanczos vectors have
been calculated, effectively restarting the procedure. The effect on the
resulting spectrum if more Lanczos vectors are used is very small.

\begin{figure}[t]
  \begin{center}
    \epsfig{file={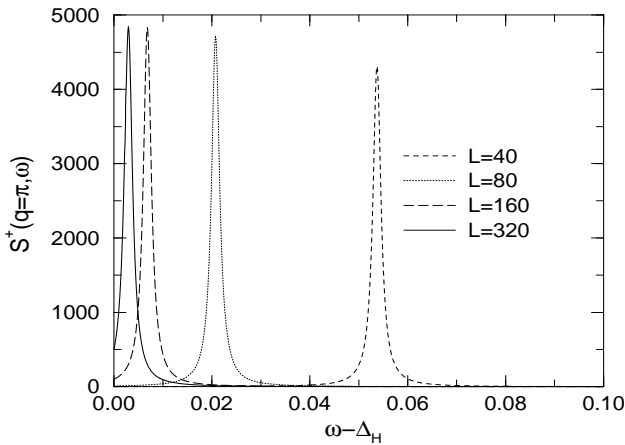}, height=\figheight,width=\figwidth, angle=-90}
    \caption
    {
	The single magnon excitation at $q=\pi$ in spin-1 chains of different length 
	versus the frequency minus the Haldane gap $\Delta_H$.
	Calculations with $m=128$ states kept and broadening factor 
	$\eta=0.001$
    }
  \label{SP1.q=pi.Splus}
  \end{center}
\end{figure}

Figure \ref{SP1.q=pi.Splus} shows $S^+(q,\omega)$ for $q=\pi$.
Most of the weight is in one single peak, and with increasing system size
this peak moves towards the Haldane gap.
To be sure that keeping $m=128$ states in the DMRG basis is enough,
we compare results with those from calculations with $m=64$ and $m=256$ states.
Table \ref{Tab1} shows the truncation error depending
on the system size and number of states. The truncation errors are 
relatively small even with \mbox{$m=64$} states, and the difference
between those with \mbox{$m=128$} and $m=256$ states are small. 

{
\table
\caption{ The truncation error $P(m)$
   as a function of the length $L$  of the spin-1 chain and 
   the number $m$ of states kept. 
}
\begin{tabular}{cccc}
	$L$	& 	$P(m=64)$ 	& $P(m=128)$  	& $P(m=256)$   \\
\tableline
\\
40 & $1.1 \times 10^{-6}$ & $2.1 \times 10^{-8}$ & $1.2 \times 10^{-8}$\\
80 & $1.6 \times 10^{-6}$ & $3.0 \times 10^{-8}$ & $3.5 \times 10^{-9}$\\
160& $1.7 \times 10^{-6}$ & $3.2 \times 10^{-8}$ & $8.6 \times 10^{-9}$\\
320& $1.5 \times 10^{-6}$ & $3.2 \times 10^{-8}$ & $3.0 \times 10^{-9}$\\
\end{tabular}
\label{Tab1}
}

\vspace{3mm}

\begin{figure}[t]
  \begin{center}
    \epsfig{file={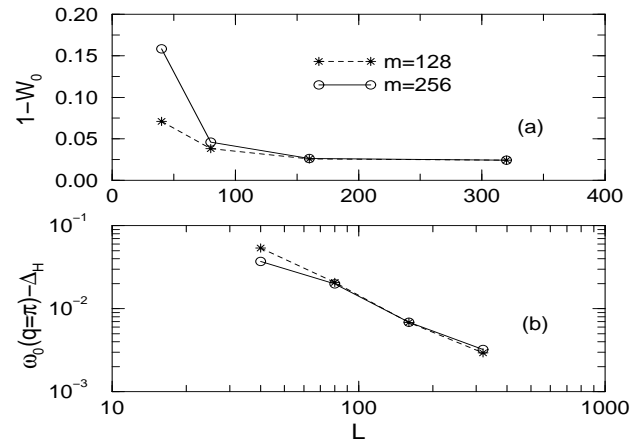}, height=\figheight,width=\figwidth, angle=-90}
    \caption
    {
	The upper plot ($a$) shows the weight of the first peak $W_0$ 
	versus the system size $L$. The lower plot ($b$) shows the difference 
	of the frequency of first peak $\omega_0$ 
	and the Haldane gap $\Delta_H$. The difference tends to zero as
	the system size increases.
    }
  \label{SP1.weight_and_omega.q=pi}
\end{center}
\end{figure}

Table \ref{Tab2} and Fig.\ref{SP1.weight_and_omega.q=pi}(a) show 
the weight in the first peak. The weight grows with the system size, but seems
to go to $W_0=97.6\%\pm0.1\%$ instead of $100\%$. This value is important for 
estimating how good a single mode approximation for this excitation is. 

{\table
\caption{ The weight $W_0$ of the first peak 
	as a function of the length $L$ of the spin-1 chain and number $m$ of 
	states kept. 
}
\begin{tabular}{cccc}
	$L$	& 	$W_0(m=64)$	& $W_0(m=128)$	& $W_0(m=256)$ \\
	\tableline
	40	& 	0.9346		& 0.9288 	& 0.8415\\
	80	&	0.9675		& 0.9617	& 0.9540\\
	160	&	0.9753		& 0.9742	& 0.9737\\
	320	&	0.9755		& 0.9755 	& 0.9756 \\
\end{tabular}
\label{Tab2}
}

\vspace{3mm}

\newpage

Table \ref{Tab3} and Fig. \ref{SP1.weight_and_omega.q=pi}(b) show the
position of the first peak $\omega_0$. We expect this value to converge
against the
Haldane gap $\Delta_H=0.41050(2)$ for big systems. For the $L=320$ site
chains our result is only $0.7\%$(128 states) above this value, and 
Fig. \ref{SP1.weight_and_omega.q=pi}(b) indicates that the distance between
the Haldane gap and the peak vanishes for big systems.

\begin{figure}[t]
  \begin{center}
    \epsfig{file={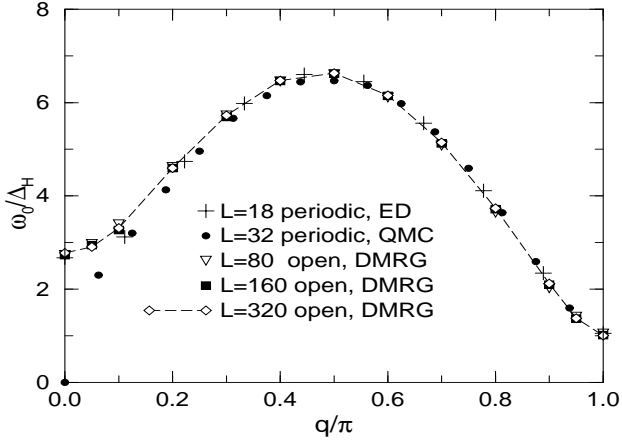}, height=\figheight,width=\figwidth, angle=-90}
    \caption
    {
	The single magnon line of the Spin 1 Heisenberg antiferromagnet.
	DMRG results with $m=128$ states per block, and 
	three Lanczos vectors as target states.
	Quantum Monte Carlo(QMC)\cite{Takahashi:1989} and exact 
	diagonalization\cite{Takahashi:1993}(ED) data are from the
	work of Takahashi.
    }
	\label{SP1.SingleMagnonLine}
  \end{center}
\end{figure}

By determining the frequency at which the first peak is for different 
wavevectors, we can calculate the single magnon dispersion relation. 
Figure \ref{SP1.SingleMagnonLine} shows the dispersion relation determined
with the Lanczos vector method, as well as quantum Monte 
Carlo\cite{Takahashi:1989} and exact diagonalization\cite{Takahashi:1993}.
The different sets of data are in good agreement, except for small $q$,
where the quantum Monte Carlo results are smaller than ours.

In summary, we have shown that the Lanczos vector method works very well for
the dynamical spectrum of the antiferromagnetic spin-1 Heisenberg model.

{\table
\caption{ The position of the first peak $\omega_0$
	as a function of the length $L$ of the spin-1 chain and number $m$ of 
	states kept. 
}
\begin{tabular}{cccc}
$L$	& 	$\omega_0(m=64)$& $\omega_0(m=128)$& $\omega_0(m=256)$ \\
\tableline
40	& 	0.4657		& 0.4642 	& 0.4477 \\
80	&	0.4326		& 0.4313	& 0.4304 \\
160	&	0.4192		& 0.4173	& 0.4174 \\
320	&	0.4159		& 0.4134	& 0.4137 \\
\end{tabular}
\label{Tab3}
}

\vspace{3mm}

\section{The spin-1/2 Heisenberg model}
\label{Spin1d2Sect}
As we have already pointed out, the spin-1 chain is a relatively easy case.
Since there is an excitation gap and the correlation length is short, the 
ground state properties can be calculated with high accuracy with
a relatively small DMRG basis. Following the example of 
Hallberg\cite{Hallberg:1995}, we now investigate the antiferromagnetic
spin-1/2 chain. This is a more difficult case, since it has no excitation
gap and a diverging correlation length, requiring a slightly bigger DMRG
basis. More importantly, instead of a single peak with most of the weight
in it, for $S^+(\omega,q)$ an excitation band is found.
It has a lower 
boundary\cite{Cloiseaux:1962}
\begin{equation}
	\omega^l_q = \frac{\pi}{2} \mid \sin(q) \mid \;,
	\label{SP1d2.lower_edge.eq}
\end{equation}
and upper boundary\cite{Yamada:1969} 
\begin{equation}
	\omega^u_q = \pi \mid \sin(q/2) \mid \;.
	\label{SP1d2.upper_edge.eq}
\end{equation}

For this band the following structure was proposed\cite{Mueller:1981}:
\begin{equation}
	S^{zz}(q,\omega)=\frac{A}{\sqrt{\omega^2-{\omega^l_q}^2}}\Theta(
\omega-\omega^l_q)\Theta(\omega^u_q-\omega) \;.
\label{bandstruct.eq}
\end{equation}

In a finite chain there can be no continuous energy band. Instead, we expect
separate peaks in the same region, and that the number of peaks increases 
as the system size is increased. 

\begin{figure}[t]
  \begin{center}
    \epsfig{file={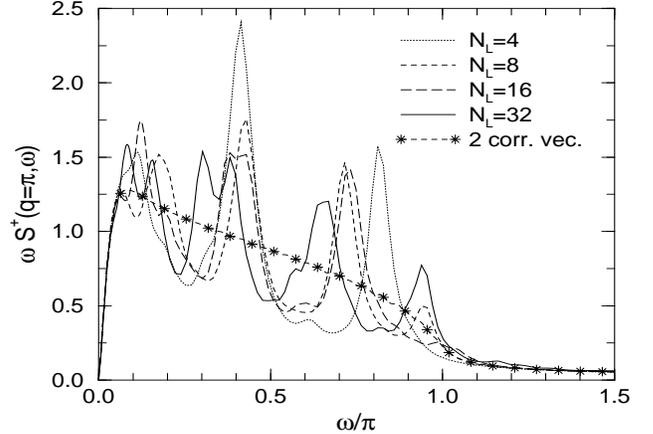},
	 height=\figheight,width=\figwidth, angle=-90}
    \caption
    {
	The spectral weight function at $q=\pi$ in a 160 site spin-1/2 chain
	with $\eta=0.1$
	for calculations with different numbers of Lanczos target state $N_L$
	and with the correction vector method.
	In all calculations $m=256$ states were kept in the DMRG basis.
	The correction vector method works well, while the Lanczos method
	is far from convergence.
    }
    \label{SP1d2.L=160.q=pi.SpectralWeight.eta=0.1}
  \end{center}
\end{figure}

This poses a problem to the Lanczos vector method. Figure 
\ref{weight_vs_Lanczosvector.Spin1_and_1d2} shows that the weight of 
the Lanczos vectors in the spectral weight function decreases only very 
slowly. This means that a lot of Lanczos vectors are needed to describe the 
correlation function well. 
Figure \ref{SP1d2.L=160.q=pi.SpectralWeight.eta=0.1} shows 
the spectral weight at $q=\pi$ determined with different numbers of Lanczos
vectors. For reference we also show results using the much more accurate
correction vector method, discussed in the next section.
The spectral weight function strongly depends on the number of
Lanczos target states, and even for 16 and 32 Lanczos vector target states
there is no sign of convergence. 
Bigger numbers of target states $N_L$ would result
in very long calculation times, and would also require 
that we keep substantially more states per block.

While the high energy part of the spectral weight function is not accessible
with the Lanczos vector method, the position of the first peak is only
weakly dependent on the number of Lanczos vector target states $N_L$.
Table \ref{Tab4} shows the position of the first peak $\omega_0$
in a $L=160$ site chain at $q=\pi$ calculated with $m=256$ states kept. 
The dependence of the position of the first peak on the number of
Lanczos vector target states is small. The reason why it is shifted
to higher values if more target states are used are the increasing
truncation errors arising from targeting more states with fixed $m=256$.

{\table
\caption{
The position of the first peak $\omega_0$ depending on the number of Lanczos
vector target states $N_L$ in the spectral weight function of
an antiferromagnetic $L=160$ site spin-1/2 chain at $q=\pi$. $m=256$ states 
were kept in the DMRG basis.
}
\begin{tabular}{ccc}
	$N_L$ &  $\omega_0$ & $P$\\
	\tableline
	4	& 0.0269 & $9.4\times10^{-8}$	\\
	8	& 0.0297 & $6.2\times10^{-6}$	\\
	16	& 0.0380 & $1.5\times10^{-4}$	\\
	32	& 0.0644 & $6.9\times10^{-4}$
\end{tabular}
\label{Tab4}
}

\vspace{3mm}

\begin{figure}[t]
  \begin{center}
    \epsfig{file={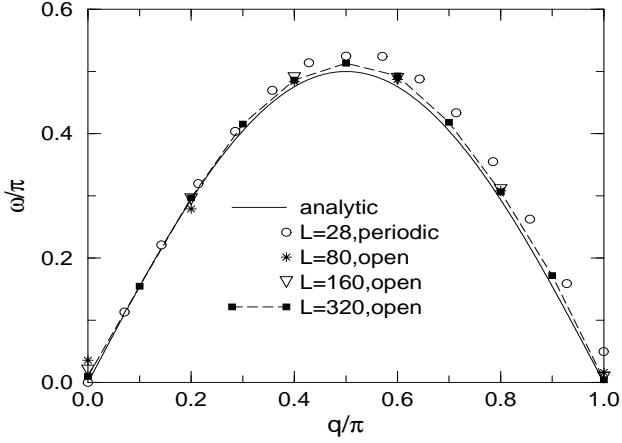}, height=\figheight,width=\figwidth, angle=-90}
    \caption
    {
	The lower bound of the excitation band in the Spin 1/2 Heisenberg 
	model. The results for the 28 site periodic system are from 
	Hallberg\cite{Hallberg:1995}.
    }
  \label{SP1d2.lower_bound_vs_q}
  \end{center}
\end{figure}

Keeping $m=256$ states in the DMRG basis, and targeting four Lanczos vectors, 
we have determined the lower edge of the energy band for different $q$. Figure 
\ref{SP1d2.lower_bound_vs_q} shows the dispersion relation. For the longer 
systems it compares increasingly well with the analytic result 
(Eq. \ref{SP1d2.lower_edge.eq}).

\section{Correction vectors}
\label{CorrVecSect}
In the previous section we found that the Lanczos vector method works very 
well if only
low-energy properties of the dynamical correlation function are of interest, 
but that it is unable to reproduce higher energy properties like the
shape of the excitation band in the spin-1/2 model.
Instead of using the Lanczos vector method, the spectrum can be
calculated directly for a given $z=\omega + i \eta$ by using a correction
vector. To do this, the following states must be included as target states:
\begin{center}
\begin{tabular}{lll}
		$|0 \rangle$ & &	the ground state\\
		$|A_q \rangle$ & $= A_q \,| 0 \rangle$ & 	the first Lanczos vector\\
		$|x(z) \rangle$ &$=
	\frac{1}{z-H} \, {|A_q \rangle}$ &  the correction vector
\end{tabular}
\end{center}

Since a finite broadening factor $\eta$ is used, the correction
vector becomes complex. To avoid the use of complex numbers, we split
the correction vector into real and imaginary part, both used as target states:
\begin{equation}
	| x(z) \rangle = {| x^r(z) \rangle} + i \, {|x^i(z) \rangle} \;.
\end{equation}

The equation for the correction vector is split into real 
and imaginary parts $|x^r(z) \rangle$ and $|x^i(z) \rangle$, and the
imaginary part is found by solving
\begin{equation}
	\left( (H-\omega)^2 + \eta^2 \right) \,{| x^i(z) \rangle} =
	-\eta \, {| A_q \rangle}
	\label{ImCorrVecEq}	
\end{equation}
using the conjugate gradient method. 
Note that this equation system gets more singular as $\omega$ gets closer
to an eigenenergy of the Hamiltonian, and as the broadening factor $\eta$ gets
smaller. For large $\eta$ the Hamiltonian is close to a diagonal matrix,
and the conjugate gradient method converges much faster than for small $\eta$.
This means that a large $\eta$ results in short calculation times, but it 
also limits the resolution of the spectrum.
The convergence is also slowed down because energy gaps in $H-\omega$ are 
squared in Eq. (\ref{ImCorrVecEq}), and the convergence rate of the conjugate
gradient method depends on the gap between the lowest and the next lowest 
eigenvector.

The real part of the correction vector is calculated directly:
\begin{equation}
	| x^r(z) \rangle = \frac{1}{\eta} (\omega-H) \,{| x^i(z) \rangle};.
\end{equation}

Using the correction vector, the Green's function can be calculated directly:
\begin{equation}
	G(q,z) = \langle A_q  \mid x(z) \rangle
\end{equation}
Taking these states ($|0\rangle$,$|A_q\rangle$ and $|x(z) \rangle$)
as target states and optimizing the DMRG basis to represent them allows for
a very precise calculation of the Green's function for
a given frequency $\omega$ and broadening factor $\eta$. 
Unfortunately, the correction vector has to be calculated separately
for every $\omega$.
If the correction vector does not change very rapidly with
$\omega$, the DMRG basis that is optimized to represent the correction 
vector at a certain $\omega$, should
also be able to represent correction vectors for nearby frequencies.

Although the correction vectors are needed as target states to determine
the DMRG basis, it is more efficient to use the Lanczos vector
method to determine the complete spectra within that basis.
Using the correction vector, but no Lanczos vectors except the first one, as
a target state, DMRG sweeps are performed until the basis is converged 
(typically two or three sweeps). 
The dynamical correlation function is then calculated in the same way as
in the Lanczos vector method: when the left and right block have the same 
size Lanczos vectors are calculated until orthogonality is lost.
Since this method is almost exact in the given basis, in principle it should
yield the same result as using the correction vector method in the same 
basis, and it is much faster. Of course, while the spectrum is produced
for all $\omega$, it is only accurate near the $\omega$ used to
produce the correction vector.

\begin{figure}[t]
  \begin{center}
    \epsfig{file={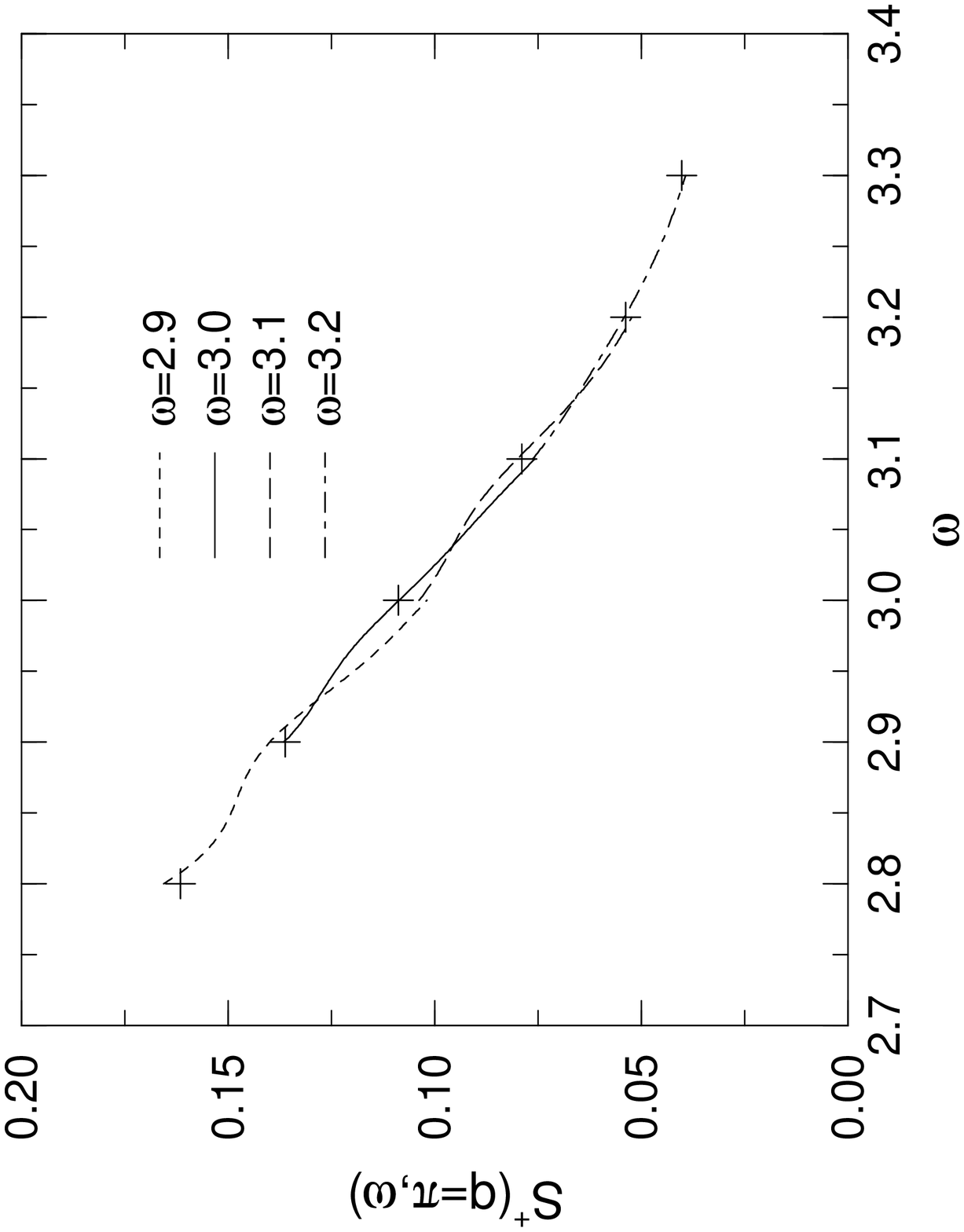},
	 height=\figheight,width=\figwidth, angle=-90}
    \caption
    {
	The spectral weight in a 80 site spin-1/2 chain at $q=\pi$
	with one correction vector as a target
	state and 128 states in the DMRG basis and $\eta=0.1$.
	The crosses show the results calculated directly from the
	correction vectors, the lines show the parts calculated with
	the Lanczos procedure in the basis optimized for the correction
	vectors. For example, the dashed line from $\omega=2.8$ to $\omega=3.0$
	is calculated in the basis that is optimized for the correction
	vector with $\omega=2.9$.
    }
   \label{SP1d2.correctionvector.1om.eta=0.1.128states}

   \epsfig{file={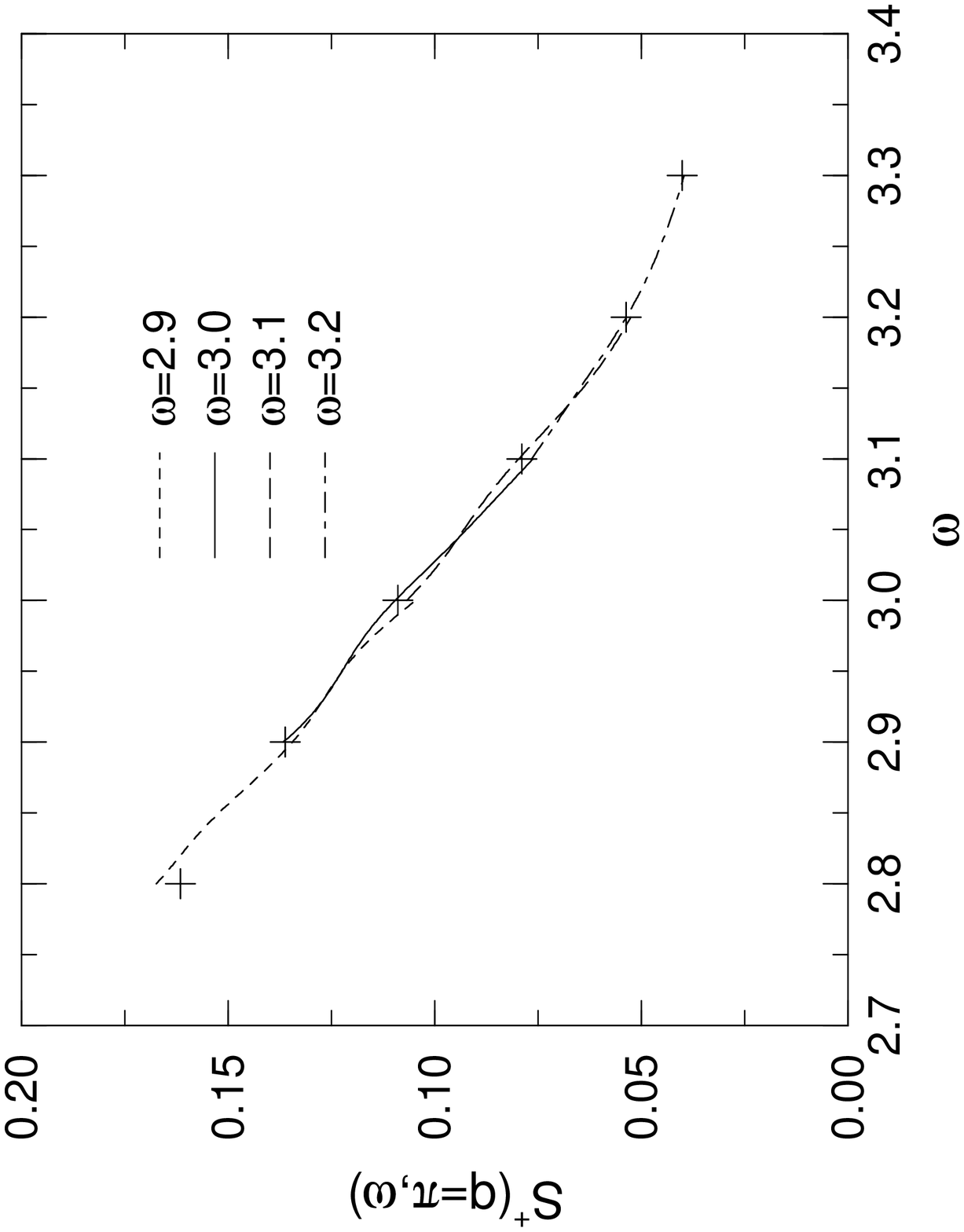},
	 height=\figheight,width=\figwidth, angle=-90}
   \caption
    {
	The spectral weight in a 80 site spin-1/2 chain  at $q=\pi$
	with one correction vector as a target
	state and 256 states in the DMRG basis and $\eta=0.1$.
    }
   \label{SP1d2.correctionvector.1om.eta=0.1.256states}
   \end{center}
\end{figure}

To determine the range of $\omega$ in which the truncated basis is good enough to 
calculate the spectral function, we compare calculations with different 
frequencies for the correction vectors. Figure 
\ref{SP1d2.correctionvector.1om.eta=0.1.128states} shows the spectral
weight function near the upper edge of the excitation band in the 
antiferromagnetic spin-1/2 chain. This is an especially difficult region, 
because there are a lot of peaks at lower energies.
We did several calculations with correction vectors for different $\omega$,
calculating the spectral weight in the region around the targeted
frequencies. With $m=128$ states kept, the overall shape reproduces the
edge of the band accurately. Deviations between the pieces of the spectrum
calculated with different correction vectors can be seen. The small mismatch
between the very accurate values calculated directly from the correction 
vectors, and the corresponding part obtained with the Lanczos vector method
are due to numerical errors in the Lanczos procedure. 
By keeping more states this result can be improved. 
Figure \ref{SP1d2.correctionvector.1om.eta=0.1.256states} shows
the same part of the spectrum with $m=256$ states kept. Now the different
parts match very well. Further improvement would be possible by keeping more
states or reducing the distance between the frequencies of the correction
vectors.

\begin{figure}[t]
  \begin{center}
    \epsfig{file={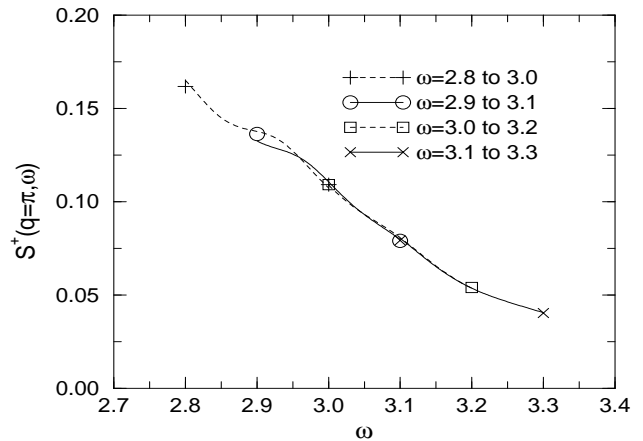},
	 height=\figheight,width=\figwidth, angle=-90}
    \caption
    {
	The spectral weight function in a 80 site spin-1/2 chain at $q=\pi$,
	with 128 states in the DMRG basis and $\eta=0.1$. Two correction
	vectors are used as target states, and the plot shows the values
	calculated directly with these correction vectors, and the connecting
	lines show the interpolated spectral weight calculated with the Lanczos
	procedure in the basis optimized for the correction vectors.
    }
	\label{SP1d2.correctionvector.2om.eta=0.1.128states}
  \end{center}
\end{figure}

\begin{figure}[t]
  \begin{center}
    \epsfig{file={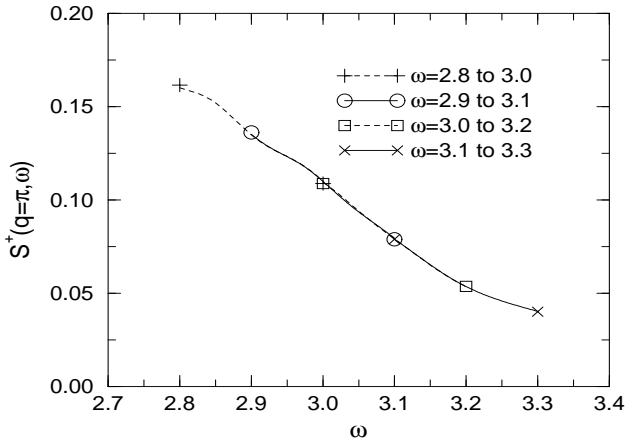},
	 height=\figheight,width=\figwidth, angle=-90}
    \caption
    {
	The spectral weight function in a 80 site spin-1/2 chain at $q=\pi$,
	with  256 states in the DMRG basis and $\eta=0.1$. Two correction
	vectors are used as target states, and the plot shows the values
	calculated directly with these correction vectors, and the connecting
	lines show the interpolated spectral weight calculated with the Lanczos
	procedure in the basis optimized for the correction vectors.
   }
    \label{SP1d2.correctionvector.2om.eta=0.1.256states}
  \end{center}
\end{figure}

This result can be improved even further. Instead of using just one 
correction vector as a target state, we try using
two correction vectors at the same time. The spectral weight can then
be interpolated 
for frequencies between these two correction vectors. With a broadening
factor of $\eta=0.1$, a distance of $\Delta\omega=0.2$ between the two 
correction vectors seems appropriate. In Fig. 
\ref{SP1d2.correctionvector.2om.eta=0.1.128states} and 
\ref{SP1d2.correctionvector.2om.eta=0.1.256states} the results from calculations
with different frequencies for the correction vectors are plotted. 
With $m=128$ states the parts for $3.0 \leq \omega \leq 3.3$ 
match perfectly, and for $2.8 \leq \omega \leq 3.0$ they still match
better than with only one correction vector as a target state 
(Fig. \ref{SP1d2.correctionvector.1om.eta=0.1.128states}).
With $m=256$ states (Fig. \ref{SP1d2.correctionvector.2om.eta=0.1.256states})
all pieces of the spectrum match up almost perfectly. This gives us a 
consistent method to verify how good our numerical results are, and 
we find that it is possible to achieve very high accuracy.

\begin{figure}[t]
  \begin{center}
    \epsfig{file={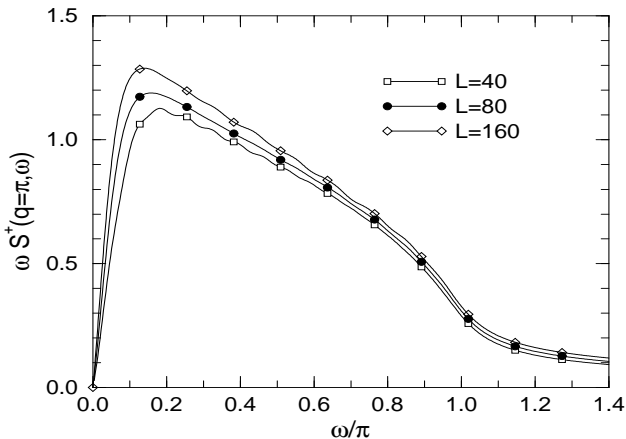},
	 height=\figheight,width=\figwidth, angle=-90}
    \caption
    {
	The spectral weight function at $q=\pi$, with different system sizes.
	The spectra were calculated with two correction vector target states
	with distances of $\Delta\omega=0.4$, a broadening factor $\eta=0.2$
	and $m=256$ states.
    }
	\label{SP1d2.different_L.q=pi.SpectralWeight.eta=0.2}
  \end{center}
\end{figure}

Now we use this method to obtain the complete spectral weight in a 
160 site chain. Keeping $m=256$ states and using two correction vectors
as target states, we  start
with correction vectors for $\omega=0$ and $\omega=0.2$. After two sweeps 
through the system the DMRG basis is converged, and the spectral weight 
function is calculated for $0 \leq \omega \leq 0.2$ using the Lanczos 
procedure.
Then we target $\omega=0.2$ and $\omega=0.4$, perform two sweeps through the
system, and calculate the spectral weight for $0.2 \leq \omega \leq 0.4$. 
Continuing this procedure, we obtain the function shown in Fig. 
\ref{SP1d2.L=160.q=pi.SpectralWeight.eta=0.1}. In contrast to the results
from the Lanczos vector method, the spectral weight function calculated using
two correction vectors shows no finite-size peaks and reflects the shape expected in 
the infinite system.  Eq. (\ref{bandstruct.eq}) predicts $1/\omega$
decay from $\omega=0$ to  $\omega=\pi$, where the spectral weight drops to
zero. The band presented in Fig. \ref{SP1d2.L=160.q=pi.SpectralWeight.eta=0.1} 
shows the correct upper and lower bound, but the spectral weight
in the band decays faster than $1/\omega$. 
To verify if this is a finite-size effect, we look at chains with different 
lengths. 
Figure \ref{SP1d2.different_L.q=pi.SpectralWeight.eta=0.2} shows 
that the spectral weight decays faster than $1/\omega$ in all system,
and the decays do not become slower for longer chains. 
The only hint at finite-size effects is the different size of the
spectral weight for different system sizes.

The small oscillations in the 40 site chain are due to the limited number 
of peaks in the relatively small system, those in the 160 site chain are 
due to the large distance of $\Delta\omega=0.4$ between the correction vectors.
They do not affect our result, and could be removed 
by using pairs of correction vectors with closer frequencies or by increasing
the size of the DMRG basis.

\begin{figure}[t]
  \begin{center}
    \epsfig{file={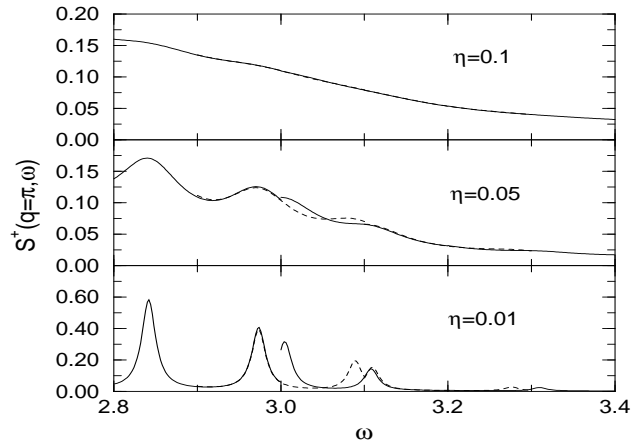},
	 height=\figheight,width=\figwidth, angle=-90}
    \caption
    {
	The spectral weight function in a 80 site chain at $q=\pi$.
	This is the same data as in 
	figure \ref{SP1d2.correctionvector.2om.eta=0.1.256states},
	with two correction vectors with a distance of $\Delta\omega=0.2$
	between them, and $\eta=0.1$. If the spectral weight function
	is plotted with a smaller $\eta$, the parts calculated with
	different frequencies for the correction vectors no longer
	match perfectly, but the peaks become easily distinguishable. 
    }
  \label{SP1d2.L=80.different_eta}
  \end{center}
\end{figure}

In Fig. \ref{SP1d2.correctionvector.2om.eta=0.1.256states} the upper
edge of the excitation band can be seen, but due to the broadening factor 
$\eta=0.1$ it is difficult to determine the position of the edge.
Since the spectral weight function is plotted by taking the peaks found
in the Lanczos procedure and plotting them with the given $\eta$, they
can as well be plotted with smaller broadening factors.
Figure \ref{SP1d2.L=80.different_eta} shows plots with $\eta=0.1$, for which
the DMRG basis was optimized, and $\eta=0.05$ and $\eta=0.01$. For the smaller
$\eta$ the different parts of the spectral weight function no longer 
match, but now the poles are easily identifiable. The position
of the peaks can also be determined by directly taking the position of 
the peaks from the Lanczos procedure. Doing this, the last peak in the band is 
found at $\omega=3.128$, compared to $\omega=\pi$ given in 
(Eq. \ref{SP1d2.upper_edge.eq}). To verify if the small difference is
a finite-size effect or a numerical error, longer chains can be studied,
or correction vectors closer to the desired region could be used. Using 
smaller $\eta$ and reducing the distance between the correction vectors
also gives higher resolution.

\section{Conclusions}
\label{ConclusionSect}

In conclusion, we have presented and tested 
two methods to calculate dynamical correlation
functions using DMRG.
We have shown that the Lanczos vector method works very well if only 
the low-energy part of the correlation function is of interest, or if
the bulk of the weight is in one single peak. We demonstrated this for
the case of the antiferromagnetic spin-1 and spin-1/2 chains, where
we could reproduce the dispersion relation of the lower edge of the
spectral weight functions correctly.

If there is an excitation band, the Lanczos vector method is unable to 
describe the higher energy part of the correlation functions. 
These parts can be determined using the correction vector method. This method
gives very precise results for frequencies at which the correction
vectors are used as target states. The Lanczos procedure
can be used in the basis optimized for the correction vectors to determine the
spectrum fast and efficiently not only at the frequency of the correction
vector, but also in the region around that. By comparing the plots
from calculations with different frequencies for the correction vectors
it is possible to estimate the range over which the spectral function
is determined correctly.
We find that remarkably good spectra can be determined if two correction
vectors are used
as target states, and the spectral function is calculated for the
frequencies between them. In the case of the excitation band in the spectral
weight function of the antiferromagnetic spin-1/2 chain, we used this method
and were able to study a system long enough and with sufficient accuracy that
no finite-size peaks were visible and excellent agreement with theoretical 
expectations was obtained.

Our results show that using these techniques, obtaining accurate dynamical 
spectral functions from DMRG is feasible and can be considered a standard
DMRG technique. The calculation time is longer than for ground state 
properties, but still manageable for single chain systems and probably
for ladders with a few chains.

\section{Acknowledgments}

The authors would like to thank H.~Monien, E.~Jeckelmann and S.~K.~Pati
for profitable discussions.
We are especially grateful to M. S. L. du Croo de Jongh for 
discussions and program design and development. This work was supported by 
the National Science Foundation under grant DMR98-70930 and the DAAD
``Doktorandenstipendium im Rahmen des gemeinsamen Hochschulsonderprogramms III von Bund und L\"andern''.

\end{document}